\documentclass[12pt,preprint]{aastex}

\usepackage{graphicx}
\usepackage{txfonts}

\newcommand{\va}{v_{\mathrm{A}i}}

\shorttitle{Swaying threads of a solar filament}
\shortauthors{Lin et al.}

\begin{document}


\title{Swaying threads of a solar filament}


\author{Y. Lin$^{1}$, R. Soler$^{2}$, O. Engvold$^{1}$, J.L. Ballester$^{2}$, \O. Langangen$^{1}$, R. Oliver$^{2}$, and L.H.M. Rouppe van der Voort$^{1}$}

\affil{$^{1}$Institute of Theoretical Astrophysics, University of Oslo, P.O. Box 1029, Blindern, N-0315
  Oslo, Norway}
  
\affil{$^{2}$Departament de F\'isica, Universitat de les Illes Balears, E-07122, Palma de Mallorca, Spain}



\begin{abstract}
From recent high resolution observations obtained with the Swedish 1-m Solar
Telescope in La Palma, we detect swaying motions of individual filament 
threads in the plane of the sky. The oscillatory character of these motions are comparable 
with oscillatory Doppler signals obtained from corresponding filament threads. Simultaneous recordings of motions in the line-of-sight and in the plane of the sky give information about the orientation of the oscillatory plane. These oscillations are interpreted in the context of the magnetohydrodynamic theory. Kink magnetohydrodynamic waves supported by the thread body are proposed as an explanation of the observed thread oscillations. On the basis of this interpretation and by means of seismological arguments, we give an estimation of the thread Alfv\'en speed and magnetic field strength by means of seismological arguments.

\end{abstract}





\section{Introduction}


When seen in a sub-arcsec resolution, solar prominences/filaments are resolved into numerous thread-like structures \citep[e.g.,][]{2005SoPh..226..239L}. These threads are magnetic in nature \citep{2008SoPh..250...31M}. Observations from the past 40 years reveal the
ever present small-amplitude oscillations in prominences and filaments. The observed oscillatory periods range
from less than 1 minute up to a few hours \citep[e.g.,][]{2008IAUS..247..152E}. The fact that line-of-sight oscillations are observed in prominences beyond the limb and in filaments against the disk suggests that the planes of the oscillation may acquire various orientations relative to the local solar reference system. Recent high resolution 
studies show waves propagating along individual threads both in quiescent filaments  \citep[e.g.,][]{2007SoPh..246...65L} and in active region filaments \citep{2007Sci...318.1577O}. In the present study the authors aim to combine motions of thin threads along the line-of-sight, as derived from Doppler data, and associated motions in the plane of the sky, which will yield information on the orientation of the plane of oscillation. 

Magnetohydrodynamic (MHD) waves have been extensively investigated and proposed as reasonable candidates to explain detected periodicities in prominences and filaments \citep[the reader is referred to reviews by][]{2002SoPh..206...45O, 2006RSPTA.364..405B}. By modeling a filament thread as a straight magnetic flux tube, partially filled with cool filament-like plasma and embedded in a much rarer and hotter coronal-like environment, \citet{2002ApJ...580..550D} and \citet{2005SoPh..229...79D} studied the linear magnetoacoustic wave modes supported by such a structure in the $\beta=0$ approximation, where $\beta$ is the ratio of the gas pressure to the magnetic pressure. On the other hand, \citet{2008ApJ...684..725S} considered a more simple model made of a homogeneous thread but took the $\beta \neq 0$ case into account and included additional effects such as nonadiabatic mechanisms and mass flow. Despite these differences in the modeling, all these works similarly conclude that, in the case of thin threads, the only trapped wave mode that is able to produce a significant nonaxisymmetric, transverse displacement of the flux tube is the so-called kink mode \citep{1983SoPh...88..179E}. Recently, \citet{2008ApJ...678L.153T} interpreted the oscillation of prominence threads reported by \citet{2007Sci...318.1577O} as kink modes. The observations of filament thread oscillations from the H$\alpha$ sequences reported in the present paper seem to be also consistent with a kink wave interpretation. Here, this possibility is explored and the implications of the observations in the context of the MHD theory are discussed.

This paper is organized as follows. Section~\ref{sec_data} describes the observations and the data reduction. The observational results are presented in Section~\ref{sec_results}. Later, Section~\ref{sec_theory} contains a theoretical interpretation of the observed oscillations. Finally, the conclusions are given in Section~\ref{sec_conclusions}.

\section{Observations and data reduction}
\label{sec_data}
\begin{figure}
\epsscale{0.90}
\plotone{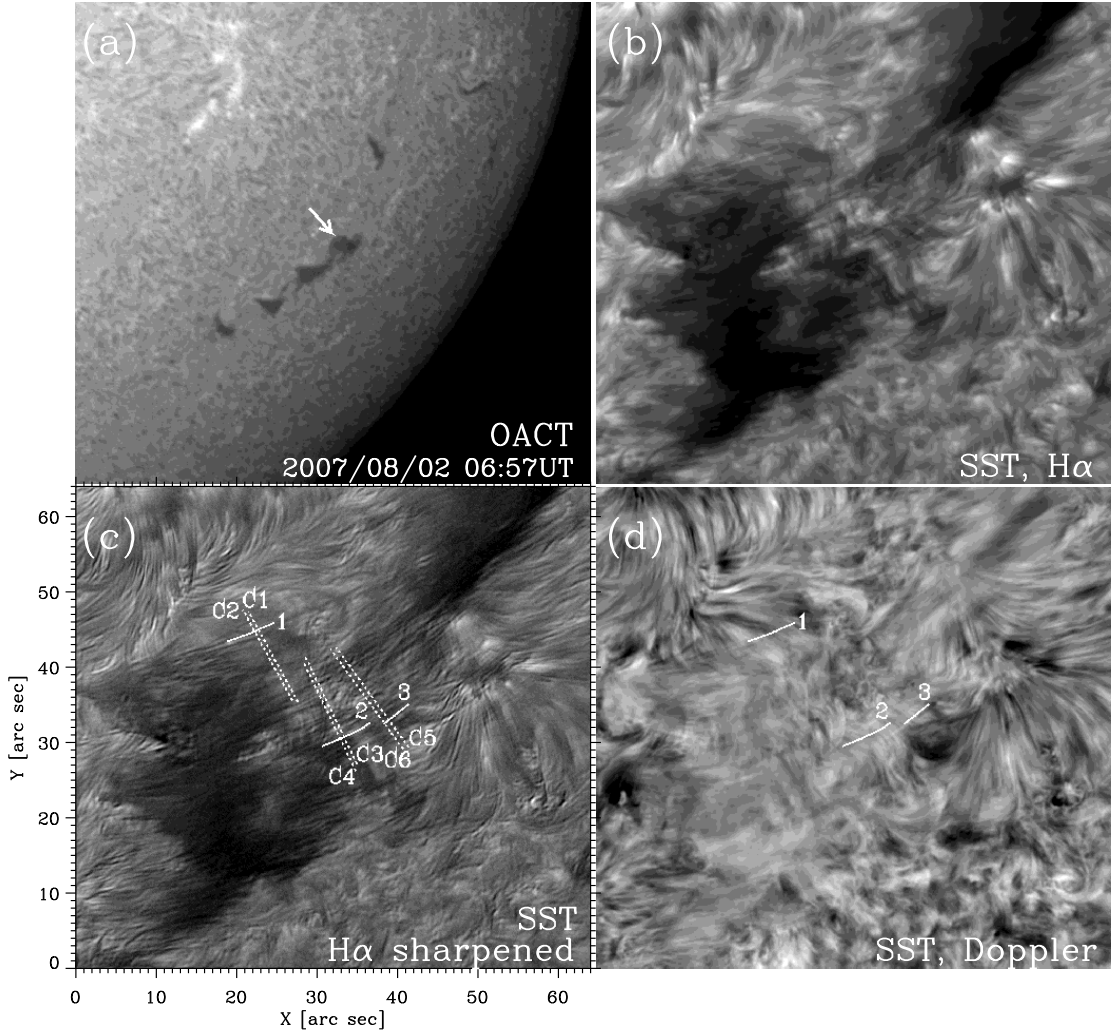}
\caption{(a): A sub-image extracted from a full-disk H$\alpha$ image from the Catania Astrophysical Observatory. The white arrow indicates the position of the SST target. Panel (b) shows one SST image in H$\alpha$ line center. Panels (c) and (d) are the corresponding sharpened H$\alpha$ image and the Dopplergram. Ten swaying threads were selected from the H$\alpha$ line center images. Three of them are marked by solid white lines in panels (c) and (d). The dashed white lines marked the locations of the cuts that were made across the swaying threads.}
\label{fig_SST_FOV_ha_dop}
\end{figure}

The target filament was observed with the Swedish 1-m Solar Telescope
(SST, \citealt{2003SPIE.4853..341S}) on August 2, 2007, at solar coordinates
S26W50 (see panel (a) of Figure~\ref{fig_SST_FOV_ha_dop}). At this heliocentric location of the target filament we observe the local horizontal and vertical directions under, respectively, 40$^{\circ}$ and 50$^{\circ}$ relative to line-of-sight. The observing field-of-view was
centered on a relatively open section of the filament where threads were
less densely packed and could be resolved individually (see panels (b) - (d) of Figure~\ref{fig_SST_FOV_ha_dop}).

The H$\alpha$ narrow-band filtergrams were obtained from the Solar Optical Universal Polarimeter (SOUP, \citealt{1981OptEng815T}) tunable filter (FWHM 12.8 pm) of the Lockheed Martin Solar and Astrophysics Laboratory. After the SOUP prefilter (FWHM 8\AA) and before the SOUP filter, 10\% of the light was reflected by a beamsplitter to a phase-diversity camera pair. These cameras provided H$\alpha$ broadband images which were used as an anchor channel in the post data reduction. The two phase-diversity cameras and the SOUP camera are Sarnoff 1K$\times$1K CCD cameras, which cover an 65$\times$65 arc\,sec field of view. The three cameras were simultaneously exposed by means of an optical chopper. The fast Sarnoff cameras allow us to record 37 images per second at one line position. However, the time for tuning line positions in the SOUP is relatively long (about 8.4 seconds). We therefore used two observing programs: (i) From 07:52UT to 08:30UT, the SOUP filter was alternating successively between three wavelength positions in the order of: H$\alpha$ red wing (+0.3\AA), line core (6562.8\AA), blue wing (-0.3\AA), line core, red wing and etc. The cadence for each wavelength is 18.5 seconds; (ii) From 08:32UT to 08:48UT, the SOUP filter was fixed in the H$\alpha$ line center position, in order to observe filaments at high rate (cadence of 3.9 seconds).

We applied the Multi-Object Multi-Frame Blind Deconvolution (MOMFBD;
\citealt{2005SoPh..228..191V}) image restoration technique to the two
phase-diversity images and the SOUP images altogether for each SOUP
cycle.  Precise alignment of the sequentially recorded SOUP images is
guaranteed through the phase-diversity channel in the MOMFBD
restoration process, see e.g., \citet{2008ApJ...673.1201L} for details on
the processing of a similar data set.  This way, H$\alpha$ Dopplergrams
can be derived that are virtually free from misalignment errors due to
seeing effects.  The Dopplergrams were derived by fitting H$\alpha$ line
core and two wing intensities to a Gaussian profile.

\citet{1964PhDT........83B} introduced the so-called cloud model to account for the influence of a chromospheric line on the line shift produced by a separate overlying solar structure (``cloud"). A number of studies of this effect followed \citep[cf.][]{1994SoPh..151...75M}. In a recent quantitative study Wiik and Engvold (2009, in preparation) simulate the impact of an artificial solar filament (``cloud") on an observed, high resolution chromospheric H$\alpha$ line spectrum. The authors vary the parameters of the artificial filament, i.e. the optical thickness ($\tau$), line source function (S$_{\lambda}$), Doppler width of the line absorption ($\Delta \lambda _{D}$) and line-of-sight velocity (V$_{los}$). The study shows that unless the optical thickness $\tau \gg $1, the resulting line shift in the combined line profile is substantially less than what corresponds to the input line-of-sight velocity. In all cases of the investigated parameter ranges, for optical thicknesses in the range 0.5 $-$ 1.0, the line shift derived from the resulting H$\alpha$ line profiles yield generally too low line shift by about a factor 2. In the following we apply this correction factor to the derived line-of-sight velocities.

We obtained sets of images in time series: H$\alpha$ line core, red and blue
wings with H$\alpha$ broadband images. By using the broadband images as
reference, the restored H$\alpha$ wing images and the line core images were
aligned and de-stretched. The alignment between sequential images is
to sub-pixel accuracy (i.e., better than 0.063 arcsec). We further applied a Laplacian kernel to the H$\alpha$ images and the Dopplergrams with the aim to suppress the chromospheric background signals. Panels (b) - (d) of Figure~\ref{fig_SST_FOV_ha_dop} show one H$\alpha$ line center image and the corresponding sharpened H$\alpha$ image and the Doppler image, respectively.

\section{Observational results}
\label{sec_results}
\subsection{Swaying threads seen in H$\alpha$ images}
\begin{figure}
\epsscale{0.80}
\plotone{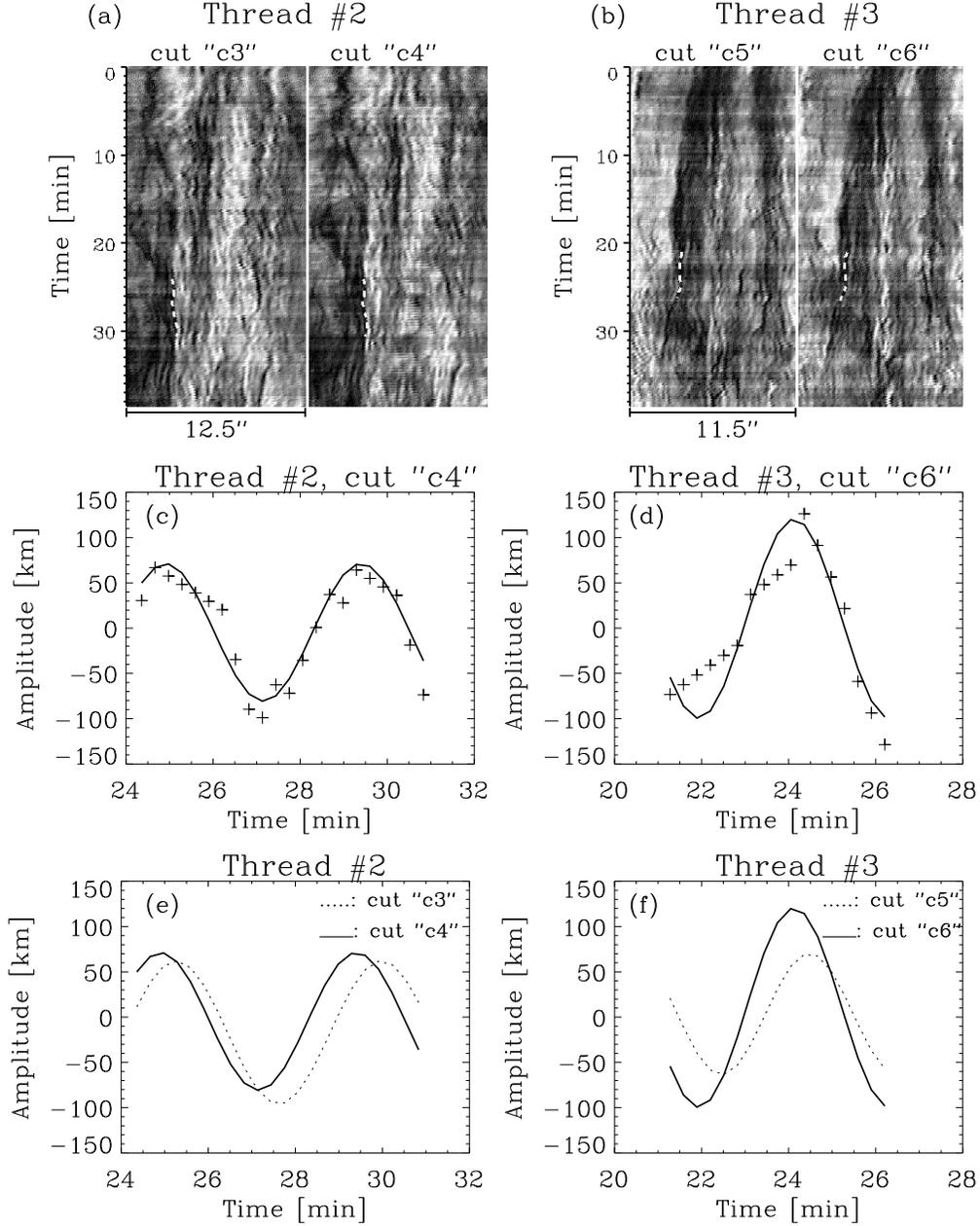}
\caption{Cuts ``c3" and ``c4" are perpendicularly crossing thread \#2. Cuts ``c5" and ``c6" are crossing thread \#3, see panel (c) of Figure~\ref{fig_SST_FOV_ha_dop}. The H$\alpha$ time-slice diagrams of these four cuts are shown in panel (a) and (b). The swaying patterns of these two threads are marked by the dashed lines. As two examples, the swaying pattern seen in cuts ``c4" and ``c6" are further shown in panel (c) and (d) as the temporal variations of the positions of the two threads ($\it plus$ $\it signs$). These variations are periodic and are fitted by sine curves ($\it solid$ $\it lines$), from which the period and the amplitude of the wave can be derived. Panel (e) and (f) show the fitted sine curves of both cuts for each thread. The phase difference between the two corresponding cuts can be used to measure the phase velocities of the waves, see text for details.}
 \label{fig_Ha_timeslice_cuts}
\end{figure}

The H$\alpha$ filtergrams resolve many thin threads in the filament. These threads are more clearly seen in the sharpened H$\alpha$ images. The H$\alpha$ sequences with cadences of 3.9\,s and 18.5\,s show that some filament threads sway back and forth in the plane of sky. Ten such swaying threads were selected for investigation. Three of them are marked as solid white lines in panel (c) and (d) of Figure~\ref{fig_SST_FOV_ha_dop}. For each thread, two or three perpendicular cuts were made to measure the properties of possible waves (see panel (c) of same figure). The H$\alpha$ time-slice diagram of each cut clearly shows the swaying signature, see panels (a)-(d) of Figure~\ref{fig_Ha_timeslice_cuts}.

Filament threads undergo a variety of changes that make them hard to follow for more than 5-10 minutes at a time. Their contrast changes and makes them appear and disappear within this time scale. They all move continuously (\citefullauthor{1990SoPh..127..109Z}, \citeyear{1990SoPh..127..109Z}; \citealt{2003SoPh..216..109L}) presumably an effect of motion of their anchoring points in the photosphere. They also often mingle with neighboring threads and observers lose track of them. However, the swaying motion is a separate, characteristic feature of the threads, which may be monitored and studied within the limited time frame.

The temporal variations of the positions of the threads are fitted by
sine curves, see panel (c)-(f) of Figure~\ref{fig_Ha_timeslice_cuts},
from which the period (P)  and the amplitude (A) of the wave are derived.
Assuming the wave is propagating along a such thread, the phase difference
between the two fitted curves represents the time interval when the wave
passes through the two cuts. We derive the time interval (T) from the maximum cross-correlation of the two curves. Taking one of
the two curves as a reference, we shift the other curve according to the
derived time interval. The deviations of corresponding points between the reference curve and the
shifted curve give the uncertainty in the measurement of time interval.
Given the distance (L) between the two cuts, one obtains the phase velocities of the waves from 
V$_{ps}$ = $\frac{\textrm L}{\textrm T}$. The uncertainties of the phase velocity can be calculated from the uncertainties in the time interval, since the distance (L) between two cuts is a constant.

Table~\ref{tab:tab_Ha_swayingthreads} lists the measured wave properties of the ten selected H$\alpha$ swaying threads. These are short period waves (mean period $\sim$ 3.6\,min) and the mean velocities of the swaying motions ($\frac{2\pi A}{P}$) are about 2\,km\,s$^{-1}$.

\begin{figure}
\epsscale{0.70}
\plotone{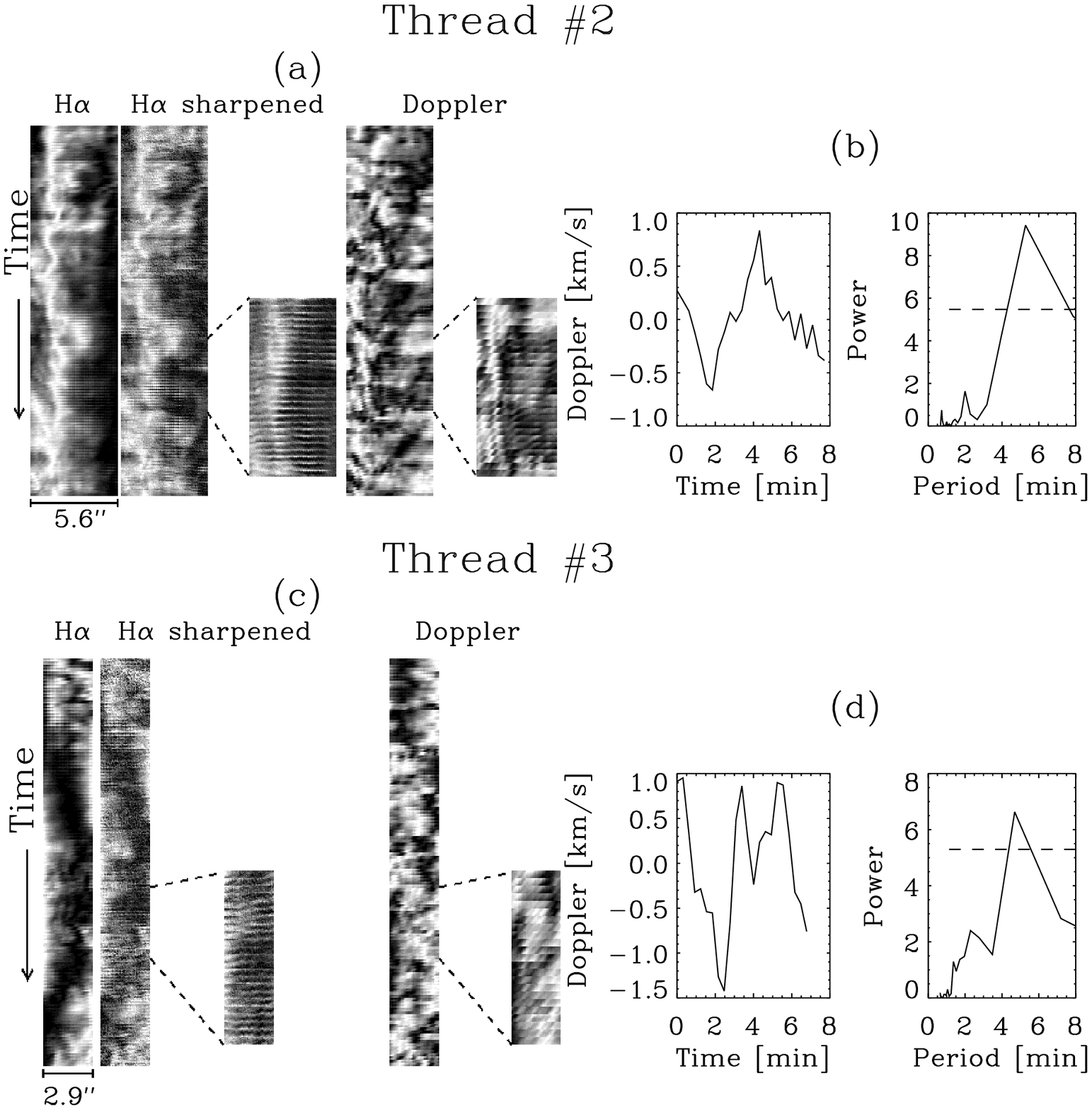}
\caption{Panels (a) and (c): The time-slice diagrams of thread \#2 and \#3 in H$\alpha$ line center, sharpened H$\alpha$ line center images and the Dopplergrams. The two short sections of time-slice diagrams in each panel correspond to the period where the swaying effects of the threads are visible in H$\alpha$ line center (c.f., Figure~\ref{fig_Ha_timeslice_cuts}). Due to the swaying effect, these two sections were reconstructed by adjusting the thread positions slightly. Panels (b) and (d): The time variation of the mean Doppler velocities averaged over each slice and the corresponding power spectrum. The dashed lines show the 90\% confidence level. The detected periods are 5.3 minutes for thread \#2 and 4.7 minutes for thread \#3.}
 \label{fig_thread2-3_dop_osci}
\end{figure}

\subsection{Thread oscillations derived from Dopplergrams}
The 40-min H$\alpha$ Dopplergrams with the cadence of 18.5\,s allow us to look for the oscillations in the line-of-sight direction. Thread \#1 $-$ \#3 are selected from this sequence. As two examples, Figure~\ref{fig_thread2-3_dop_osci} shows the time-slice diagrams in both H$\alpha$ and the Doppler for thread \#2 and \#3. 

Because the position of a slice (marked as some white solid lines in panel (c) of Figure~\ref{fig_SST_FOV_ha_dop}) is defined from one image and is fixed, when plotting a time-slice diagram, the filament thread may move out of such a fixed slice ``window" due to the swaying effect. Therefore, we reconstruct the short sections of the time-slice diagrams, where the swaying threads are seen in H$\alpha$, by adjusting the slice positions slightly (see panel (a) and (c) of Figure~\ref{fig_thread2-3_dop_osci}). The corresponding short section of the Doppler time-slice diagrams are reconstructed accordingly. 

For thread \#2 and \#3, the temporal variations of their mean Doppler velocities averaged over each slice indicate that the threads oscillate in the line-of-sight (see panels (b) and (d)  of Figure~\ref{fig_thread2-3_dop_osci}). From the power spectrum, the waves are short period ($\sim$ 5\,min). Compared with thread \#2 and \#3, the Doppler signals of thread \#1, which is not shown here, are relatively weak.

The amplitude of the derived line-of-sight velocity ranges typically between 1 and 2 km\,s$^{-1}$. For threads \#2 and \#3, they are 0.8 km\,s$^{-1}$ and 1.9 km\,s$^{-1}$, respectively. The quantitative study of Wiik and Engvold (2009, in preparation) also demonstrates that in cases when the optical thickness of the artificial filament $\tau  \approx$ 1 the velocity fluctuation caused by the fine structure of the chromosphere below is suppressed to yield an uncertainty $\sigma \approx$ 0.5 km\,s$^{-1}$. Applying this intrinsic error (0.5 km\,s$^{-1}$) caused by the chromosphere below and the ``cloud model"-correction, the two line-of-sight velocities raise to 1.6$\pm$0.5 km\,s$^{-1}$ and 3.8$\pm$0.5 km\,s$^{-1}$.

\subsection{Combination of the observed oscillations in the two orthogonal directions}

The two selected H$\alpha$ threads are found oscillating both in the plane of sky and in the line-of-sight, with a similar period. Since the location of the target filament is S26W50, the measured swaying motion and oscillation of velocity in line-of-sight represent inevitably the projected components of the oscillating motions. As noted earlier, the oscillations of the filament threads are probably polarized and the plane of oscillation may attain various orientation relative to the local reference system. Swaying motions will be most clearly observed when a thread sways in the
plane of the sky while Doppler signals will be strongest for oscillations
along the line of sight.

Combining the corresponding components derived from the swaying motion of the two threads, which are 1.8$\pm$0.2 km\,s$^{-1}$ and 2.3$\pm$0.2 km\,s$^{-1}$, and the line-of-sight oscillatory motion, 1.6$\pm$0.5 km\,s$^{-1}$ and 3.8$\pm$0.5 km\,s$^{-1}$,  we may derive the full vectors as 2.4$\pm$0.5 km\,s$^{-1}$ and 4.4$\pm$0.5 km\,s$^{-1}$ which are directed at angles 42$\pm$10$^{\circ}$  and 59$\pm$4$^{\circ}$, respectively, relative to the plane of the sky. Since the heliocentric position of the target filament is W50, the corresponding local vertical direction becomes 50$^{\circ}$ relative to line-of-sight and we conclude that these two threads oscillate in planes which are reasonably close to vertical. It is, however, not possible based on only two cases to draw any general conclusion about the orientation of planes of oscillation of filament threads.  

From Table~\ref{tab:tab_Ha_swayingthreads}, one notices that the amplitudes of the swaying motions are very small. This calls for very high spatial resolution in observations, which may explain why it has not been noticed in earlier studies. On the other hand, for some threads, the amplitudes of the waves passing through two cuts are notably different. In other words, the amplitudes of the waves are changing while they propagate along these threads. Apparent changes can be due to spatial damping of the waves in addition to the noise in the data. The damping phenomenon appears very common in solar filaments/prominences with a damping time corresponding to from 1 to 4 periods \citep[e.g.,][]{1999joso.proc..126M, 2002A&A...393..637T}. Various possible theoretical mechanisms have been proposed \citep[e.g.,][]{2001A&A...378..635T, 2002A&A...393..637T, 2008ApJ...684..725S, 2008ApJ...682L.141A}, among which the mechanism of resonant absorption seems 
to be the most efficient one for the damping of kink waves \citep[e.g.,][]{2008ApJ...682L.141A, 2009ApJ...695L.166S}. It is also expected this  mechanism to be the dominant one for the spatial damping of propagating kink waves, although a theoretical investigation on this issue is needed. Alternatively, the observed change of the wave amplitude might also be caused by a rotation of the oscillating planes. We will look for such evidence from higher temporal resolution data (cadence less than 2 sec) in a following filament seismology study.

\begin{table}[!t]
\centering

\begin{tabular}{c|cccc}\hline

Thread \# & cut \#  & P [min] & V$_{ph}$ [km\,s$^{-1}$] & A [km] \\ \hline

1 & c1  & 3.5$\pm$0.1 & 16$\pm$3 & 79$\pm$6 \\
 & c2   & 3.9$\pm$0.1 &  & 70$\pm$7  \\ \hline

2 & c3  & 4.72$\pm$0.05 & 20$\pm$6 & 79$\pm$7   \\ 
& c4   & 4.50$\pm$0.03 &  & 76$\pm$6 \\ \hline

3 &  c5 & 3.9$\pm$0.1 & 24$\pm$6 & 67$\pm$10 \\
  & c6  & 4.4$\pm$0.1 &  & 110$\pm$9 \\ \hline\hline

4 & c7 & 3.66$\pm$0.04 & 36$\pm$6 & 88$\pm$4 \\
 & c8 & 3.69$\pm$0.04 &  & 86$\pm$4\\ \hline

5 & c9  & 3.76$\pm$0.02 & 57$\pm$9 & 96$\pm$3 \\
 & c10 & 3.78$\pm$0.03 &  & 81$\pm$3  \\ \hline

6 & c11  & 2.7$\pm$0.1 & 28$\pm$12 & 57$\pm$4 \\
 & c12  & 4.0$\pm$0.1 &  & 73$\pm$5 \\ \hline

7 & c13  & 2.0$\pm$0.1 & 62$\pm$10 & 52$\pm$3\\
 &  c14  & 1.9$\pm$0.1 &  & 59$\pm$3 \\
 & c15 & 2.0$\pm$0.1 &  & 52$\pm$4\\ \hline

8 & c16  & 3.1$\pm$0.1 & 40$\pm$6 & 56$\pm$4 \\
 & c17 & 3.0$\pm$0.1 &  & 34$\pm$2\\ \hline

9 & c18 & 2.8$\pm$0.1 & 20$\pm$3 & 34$\pm$3 \\
  & c19 & 2.6$\pm$0.1 &  & 57$\pm$4 \\ \hline

10 & c20  & 5.4$\pm$0.1 & 28$\pm$9 & 88$\pm$3 \\ 
 & c21   & 5.0$\pm$0.2 &  & 58$\pm$3 \\ \hline

\end{tabular}
\caption{For the ten selected H$\alpha$ swaying threads, the periods ($P$), phase velocities (V$_{ph}$) and the amplitudes ($A$) of the possible waves were measured. Thread $\#$1$-\#$3, indicated in Figure~\ref{fig_thread2-3_dop_osci} (c), are selected from the time series with the cadence of 18.5 seconds. Thread $\#$4$-\#$10 are from the time series with the cadence of 3.9 seconds.}
\label{tab:tab_Ha_swayingthreads}
\end{table}

\section{Theoretical interpretation of the oscillations}
\label{sec_theory}

\subsection{Kink magnetohydrodynamic waves in a magnetic flux tube}

\begin{figure}[!htb]
\centering
\epsscale{0.75}
\plotone{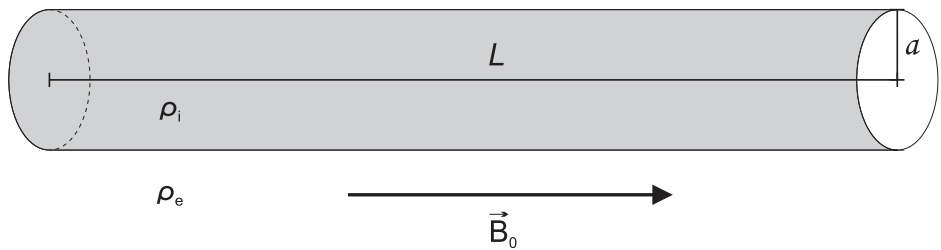}
\caption{Sketch of the model configuration. \label{fig:model}}
\end{figure}

Next, in order to perform an interpretation of the observed oscillations in terms of linear magnetohydrodynamic waves, we consider a simple model representing a prominence thread. We assume a configuration made of a straight and homogeneous cylinder of length $L$ and radius $a$, filled with filament-like plasma and embedded in an also homogeneous coronal environment. The magnetic field is taken homogeneous and orientated along the tube axis (see a scheme of the model in Fig~\ref{fig:model}). In this model, we neglect the longitudinal inhomogeneity of the thread plasma for simplicity and consider $\beta = 0$, so magnetic effects are dominant over gas pressure effects. By using cylindrical coordinates, we assume that the density profile, $\rho_0$, only depends on the radial coordinate, $r$, as follows,
\begin{equation}
 \rho_0 \left( r \right) = \left\{
\begin{array}{lll}
 \rho_i, & \textrm{if} & r \leq a, \\
\rho_e, & \textrm{if} & r > a,	
\end{array}
  \right.
\end{equation}
where $\rho_i$ and $\rho_e$ are the internal and external densities respectively. 

\begin{figure}[!t]     
\centering
\includegraphics[width=0.49\columnwidth]{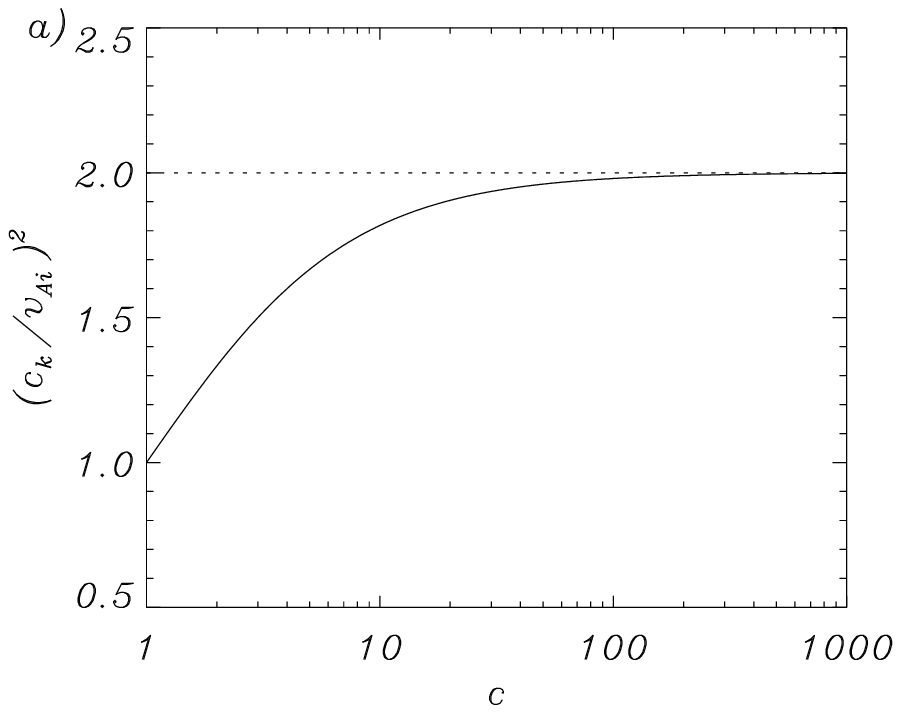}
\includegraphics[width=0.49\columnwidth]{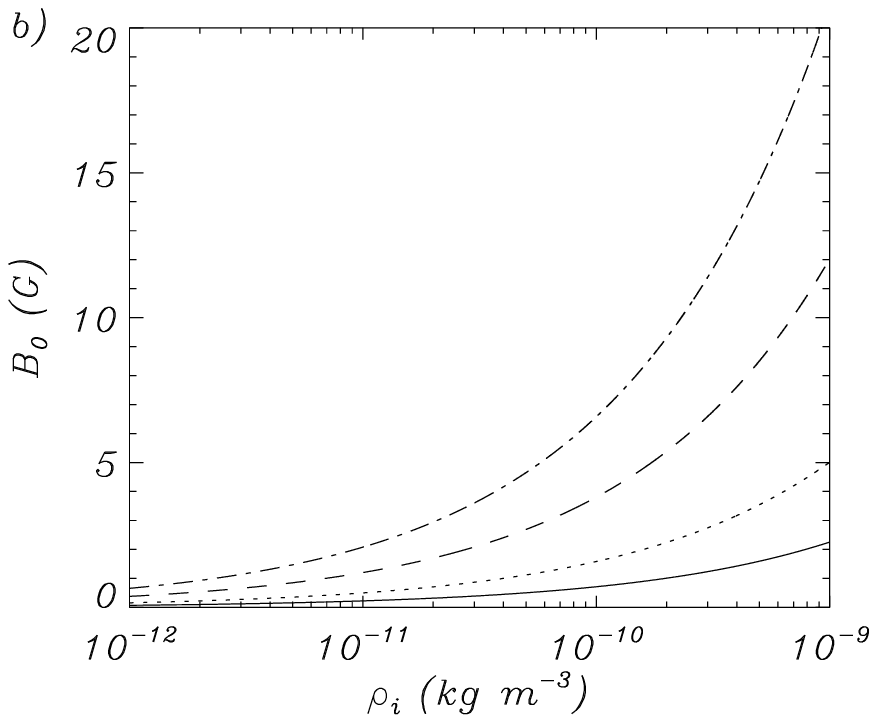}
\caption{$a)$ Ratio $c_k^2 / \va^2$ (solid line) as a function of the density contrast, $c$. The dotted line corresponds to the value of the ratio $c_k^2 / \va^2$ for $c \to \infty$. $b)$ Magnetic field strength, $B_0$, as a function of the internal density, $\rho_i$, corresponding to some selected threads: thread 1 (solid line), thread 3 (dotted line), thread 5 (dashed line), and thread 7 (dash-dotted line).}
\label{fig:theo1} 
\end{figure}

The linear magnetohydrodynamic wave modes supported by this flux tube model have been investigated in detail \citep[e.g.,][]{1982SoPh...75....3S,1983SoPh...88..179E,1986SoPh..103..277C}. Among the possible wave modes, the kink mode is the only one that can produce a significant transverse displacement of the cylinder axis, and so it is the best candidate for an interpretation of the present oscillations. Moreover, the kink mode produces short-period oscillations of the order of minutes for typical filament plasma conditions, which is also compatible with the measured periods. Since the observed threads appear to be very thin and long structures, it seems reasonable to consider the so-called thin tube approximation, i.e., $L \gg a$. In such an approximation, the kink mode phase velocity (also called the kink speed), namely $c_k$, is given by
\begin{equation}
 c_k = \sqrt{\frac{2 B_0^2}{\mu \left( \rho_i + \rho_e \right)}}, \label{eq:kinkspeed}
\end{equation}
where $B_0$ is the magnetic field strength and $\mu=4\pi \times 10^{-7} NA^{-2}$ is the magnetic permittivity. Equation~(\ref{eq:kinkspeed}) can be rewritten in terms of the internal Alfv\'en speed, $\va = B_0 / \sqrt{ \mu \rho_i }$, and the density contrast, $c = \rho_i / \rho_e$, as follows
\begin{equation}
  c_k = \va \sqrt{ \frac{2c}{c + 1} }. \label{eq:kinkspeed2}
\end{equation}
Figure~\ref{fig:theo1}a displays the ratio $c_k^2 / \va^2$ as a function of $c$. We see that for large density contrast the curve becomes flat and so the ratio $c_k^2 / \va^2$ is then almost independent of the density contrast. For typical coronal and filament densities one has $c \approx 200$. In such a situation, the factor $\sqrt{ \frac{2c}{c + 1} }  \approx \sqrt{2}$ in Equation~(\ref{eq:kinkspeed2}) and so the kink speed directly depends on the internal Alfv\'en speed,
\begin{equation}
 c_k \approx \sqrt{2} \va. \label{eq:kinkspeed3}
\end{equation}

Now, assuming that thread oscillations observed from the H$\alpha$ sequences are the result of a propagating kink mode, the measured phase velocities, $V_{ph}$, can be related to the kink speed, so $c_k = V_{ph}$. Therefore, it is possible to use Equation~(\ref{eq:kinkspeed3}) to give a seismological estimation of the thread Alfv\'en speed, 
\begin{equation}
 \va \approx V_{ph} / \sqrt{2}. \label{eq:alfven}
\end{equation}
The use of the MHD seismology to obtain information of the plasma physical conditions has been applied by some authors in the context of coronal loop oscillations \citep[e.g.,][]{2001A&A...372L..53N,2007A&A...463..333A,2008ApJ...676L..77A,2008A&A...484..851G} and prominence oscillations \citep{2008ApJ...678L.153T}. Since the density contrast in the case of coronal loops typically has a small value, the factor with the density contrast cannot be dropped from equation~(\ref{eq:kinkspeed2}) and so it is not possible to give a direct estimation of the Alfv\'en speed from the observed kink speed. However, the much larger density of filament threads allows us to provide a more accurate determination of the Alfv\'en speed.

\subsection{Determination of the Alfv\'en speed and the magnetic field strength}

Table~\ref{tab:theo} contains an estimation of the Alfv\'en speed of each swaying thread observed in the H$\alpha$ sequences by taking expression~(\ref{eq:alfven}) into account. We see that the Alfv\'en speed varies in a wide range, which suggests that physical properties (i.e., density, magnetic field strength, etc.) significantly changes in different threads of the filament. Once the Alfv\'en speed is determined, we have a relation between the magnetic field strength and the thread density, i.e., $B_0 = \va \sqrt{\mu \rho_i}$. Since the thread density is an unknown parameter, we cannot uniquely determine the magnetic field strength or vice versa. Figure~\ref{fig:theo1}b shows the dependence $B_0$ with $\rho_i$ corresponding to four selected threads. The density has been varied in a wide range, $10^{-12}$~kg~m$^{-3} < \rho_i < 10^{-9}$~kg~m$^{-3}$, and subsequently the magnetic field strength is in the range $0.1\, \textrm{G} \lesssim B_0 \lesssim 20\, \textrm{G}$. This range agrees with the usually reported values of the magnetic field strength in prominences and filaments \citep[e.g.,][]{2002SoPh..208..253P}. From Figure~\ref{fig:theo1}b it can be also observed that the larger the density, the larger the magnetic field strength, which is compatible with the idea that active region filaments have stronger magnetic fields. By assuming a typical value of $\rho_i = 5\times 10^{-11}$~kg~m$^{-3}$ for the thread density and considering the results for the ten studied threads, one obtains a magnetic field strength which varies in the range $0.5\, \textrm{G} \lesssim B_0 \lesssim 5\, \textrm{G}$, see Table~\ref{tab:theo}. Although we have to be cautious with the values obtained in this very rough estimation, the present results suggest that the magnetic field in the filament is not homogeneous and could largely vary between threads. Also, this could happen for the density, but the results are more sensible to variations of the magnetic field strength than variations of the density.

\begin{table}[!t]
\centering
\begin{tabular}{ccc}
\hline Thread \# & $\va$~[km s$^{-1}$] & $B_0$~[G]  \\
\hline 1 & 12$\pm$2 & 0.9$\pm$0.3 \\
	2 & 14$\pm$4 & 1.1$\pm$0.5 \\
	3 & 17$\pm$4 & 1.3$\pm$0.5 \\
	4 & 26$\pm$4 & 2.0$\pm$0.4 \\
	5 & 41$\pm$6 & 3.2$\pm$0.7 \\
\hline
\end{tabular}\hspace{0.5cm}
\begin{tabular}{ccc}
\hline Thread \# & $\va$~[km s$^{-1}$] & $B_0$~[G]  \\
\hline  6 & 20$\pm$8 & 1.6$\pm$0.9 \\
	7 & 44$\pm$7 & 3.5$\pm$0.8 \\
	8 & 29$\pm$4 & 2.3$\pm$0.5 \\
	9 & 14$\pm$2 & 1.1$\pm$0.2\\
	10 & 20$\pm$6 & 1.6$\pm$0.7\\
\hline
\end{tabular}
\caption{Estimation of the Alfv\'en speed, $\va$, and magnetic field strength, $B_0$, of threads listed in Table~I by considering expression~(\ref{eq:alfven}). The value of $B_0$ has been obtained by assuming $\rho_i = 5\times 10^{-11}$~kg~m$^{-3}$.}
\label{tab:theo}
\end{table}

\section{Concluding remarks}
\label{sec_conclusions}
In this study, the SST H$\alpha$ intensity images and the Dopplergrams reveal many thin threads of a quiescent filament. The H$\alpha$ line center images in time series show a number of threads swaying back and forth in the plane of sky, with a mean period of 3.6 minutes, a mean phase velocity of 33 km\,s$^{-1}$ and a mean amplitude velocity of 2 km\,s$^{-1}$. A similar swaying pattern of filament threads is also seen in a recent Hinode Ca\,II\,H observation by \citet{2007Sci...318.1577O} and \citet{2008ApJ...676L..89B}. Okamoto et al. found some horizontal threads in an active region prominence oscillating vertically in the dark plane of sky, with periods of 130$-$250\,s and amplitudes from 400 to 1800\,km. Their measured periods are consistent with the current study. However, their amplitudes appear larger which might be because their prominence target is in an active region. \citet{2008ApJ...676L..89B} also found transverse oscillations in quiescent prominences with intermediate periods of 20-40 minutes, propagating speed of 10 km\,s$^{-1}$ and amplitudes of 100$-$250 km. The longer duration of their observations (several hours) favors the detection of longer periods. It should be noted that similar swaying motions have also been observed in solar spicules by \citet{2007Sci...318.1574D}, who interpreted the transverse motions as Alfv\'enic waves.

Two filament threads in the current dataset are found both swaying in the plane of sky and oscillating in the line of sight with similar periods. Assuming they are the two projected components of the oscillating motions, the orientation of the oscillating plane can be derived. This will be further investigated in a follow-up study based on a higher temporal resolution data (cadence less than 2\,s). 

We have interpreted the observed thread oscillations in the H$\alpha$ sequences in terms of kink magnetohydrodynamic wave modes propagating along the threads. In the context of this interpretation, we have provided a determination of the Alfv\'en speed of the studied threads. Subsequently, we have assumed a typical value for the plasma density and have estimated the magnetic field strength in the filament. High resolution observations of filament threads suggest that the plasma parameters may be varying along individual threads, which may affect, somehow, the derived magnetic field strength.

\begin{acknowledgements}
Y.L. acknowledges the Norwegian Research Councial grant FRINAT 171012. R.S., R.O., and J.L.B. acknowledge the financial support received from the Spanish MICINN and FEDER funds, and the Conselleria d'Economia, Hisenda i Innovaci\'o of the CAIB under Grants No. AYA2006-07637 and PCTIB-2005GC3-03, respectively. R.S. thanks the CAIB for the fellowship and finance support for his visit in Oslo and also thanks O. Engvold and Y. Lin for their warm hospitality during his stay in Oslo. We are very grateful to the staff of the SST for their invaluable support with the observations. The Swedish 1-m Solar Telescope is operated on the island of La Palma by the Institute for Solar Physics of the Royal Swedish Academy of Sciences in the Spanish Observatorio del Roque de los Muchachos of the Instituto de Astrof{\'\i}sica de Canarias.

\end{acknowledgements}

\clearpage

\end{document}